\documentclass[twocolumn,superscriptaddress,showpacs,floatfix,letterpaper]{revtex4}
\usepackage{graphicx,amsmath}
\usepackage[totalwidth=7.2in,totalheight=9.5in]{geometry}
\begin{document}

\newcommand{\ie}{\emph{i.e.}}
\newcommand{\eg}{\emph{e.g.}}

\title{Network Growth with Feedback}
\author{Raissa M. D'Souza}
\affiliation{Department of Mechanical and Aeronautical Engineering, University of California, Davis, CA 95616}
\affiliation{Center for Computational Science and Engineering, University of California, Davis, CA 95616}
\author{Soumen Roy}
\affiliation{Department of Mechanical and Aeronautical Engineering, University of California, Davis, CA 95616}
\affiliation{Center for Computational Science and Engineering, University of California, Davis, CA 95616}

\begin{abstract}
Existing models of network growth typically have one or two parameters or strategies which are fixed for all times. We introduce a general framework where feedback on the current state of a network is used to dynamically alter the values of such parameters. A specific model is analyzed where limited resources are shared amongst arriving nodes, all vying to connect close to the root.  We show that tunable feedback leads to growth of larger, more efficient networks. Exact results show that linear scaling of resources with system size yields crossover to a trivial condensed state, which can be considerably delayed with sublinear scaling. 
\end{abstract}

\maketitle

The prevalence and importance of network structures in physical, biological and social systems is becoming widely recognized. Current research on network growth  focuses on models which reproduce aspects of real-world networks, in particular the broad range of node degrees typically observed~\cite{BarabasiAlbert99,KRRSTU,DorogMenSam00,KrapRed01,FKP02,Bhan-copying02,dsouza-pnas07}. 
These simple and elegant models have just one or two free parameters, or strategies which are specified initially and remain unaltered even as the network grows to a massive size, starting from a few seed nodes.  Yet, the functionality and performance required of a small network may be radically different from that of a large network. Thus, it is natural that the parameters of the growth strategy should change over time as the network grows.  
The mechanisms underlying these distinct growth models can be generally classified as growth via 
either preferential attachment~\cite{BarabasiAlbert99,DorogMenSam00,KrapRed01},  copying~\cite{KRRSTU,Bhan-copying02}, or optimization~\cite{FKP02,dsouza-pnas07}. 
In preferential attachment models, the extent of the preference (\ie, the connection kernel) could be altered, tuning properties of the resulting degree distribution~\cite{DorogMenSam00,KrapRedLey00}.  In copying models, the probability of successfully copying links could be changed, thus affecting degree distribution. In optimization models, the explicit parameter values of the optimization function could be altered, leading to a range of interesting behaviors~\cite{FKP02,dsouza-pnas07,RDPKCM07}.

In this rapid communication, we introduce a framework where information on the current state of a 
network provides {\em feedback} to the system allowing it to dynamically alter and 
self-tune the parameter values throughout the growth process.  It combines local 
optimization models of growth~\cite{dsouza-pnas07,RDPKCM07} with measures 
of efficient information flow in a network~\cite{Infocom03}.   We show that with feedback, 
one can grow larger and more efficient network structures in less time. 
This framework can be applied to many systems  
exhibiting a hierarchical ``chain of command" structure.  
Simple examples are business enterprises, armed forces, etc., with the ``CEO" or the commander in chief respectively being the root node of the hierarchy.  Such a structure 
has also recently been 
found in the organization of genetic 
regulatory networks~\cite{YuGerstein06}. More generally, hierarchy appears to be a central organizing principle of complex networks, providing insight into structures such as food webs, biochemical and social networks~\cite{aaron08}. 

We are interested in growth of hierarchical networks where information flow is essential to the network's function.  Two  
basic considerations are: (i) ensuring a smooth ``flow" of commands or information {\it throughout} the structure, and,  (ii) addition of new nodes subject to constraints on resources.  More explicitly, only some fraction,  $0 < c \le 1$, of existing resources can be dedicated to optimizing new growth.  The remaining portion of the system is involved with performing some task (\eg, information processing, regulation, transport and routing), crucial to the sustenance and function of the organization. 
We show herein that  how the resources allocated for growth scale with system size $N$ 
directly impacts the resulting network structure.  Moreover, we show that incorporating feedback leads to flatter hierarchies on which information flows more efficiently, providing a quantitative underpinning to previous case studies of individual organizations where this is found in practice~\cite{biz-models}.

We consider a simple growth model incorporating (i) and (ii). 
It is a discrete time  process starting from a single root node. Let $G(t)$ denote the network at time $t$ and $N(t)$ the number of nodes.  At each time-step, an integer number of new nodes, $\lambda(t) \ge 1$,  arrive and must connect to the existing network.  In accord with (ii), the fraction,  $0 < c \le 1$, of  
resources dedicated to optimizing new growth must be shared equally by all $\lambda(t)$ arriving nodes.  Thus, each arriving node sees only 
$k(t) = [c/\lambda(t)] [N(t)]^\alpha$ 
randomly chosen {\it candidate} parent nodes, where $\alpha$ determines the scaling 
of resources and system size, \eg, $\alpha=1$ is linear scaling.  
It then chooses the one candidate parent which is optimal in some sense (with degeneracy broken by a random choice) and connects to it. Even the simple criteria, 
that optimal is the candidate closest to the root node will demonstrate the importance of feedback.  

To elaborate on (i), the efficiency of information flow on $G(t)$ quantifies the network fitness, 
${\cal F}(G(t))$, and is measured by the characteristic time-scale, $\tau_c$, (see below and~\cite{Infocom03}), 
for a weighted random walk on $G(t)$. Other measures exist, \eg, \cite{GuptaKumar}, yet $\tau_c$ is used herein due to its simplicity. 
${\cal F}(G(t))$ is assessed every $\delta$ time-steps.  If it 
is found to increase, the system is rewarded by increased arrival rate $\lambda$.  If it  
decreases, the system is penalized by decreased  
$\lambda$.
Thus, starting from initial value 
$\lambda_0 \ge 1$, due to {\em feedback}, $\lambda(t) \equiv \lambda_t$ evolves as:
    \begin{equation}
    \lambda_{t+1} = 
    \begin{cases}
    \lambda_t & \mbox{if } {\cal F}(G(t)) = {\cal F}(G(t-\delta))\\
    \lambda_t+1 & \mbox{if } {\cal F}(G(t)) > {\cal F}(G(t-\delta))\\ 
    \max[\lambda_t-1,1] & \mbox{if } {\cal F}(G(t)) < {\cal
    F}(G(t-\delta))
    \end{cases}
    \label{eqn:lambda}
    \end{equation}
$\lambda$ is thus a tunable parameter and our goal is to use feedback to tune $\lambda$ during the growth process to build larger and more efficient structures.

We capture a basic feedback loop (Fig.~\ref{fig:FBnet}~(b), inset). For a given $c$, as $\lambda$ increases, $c/\lambda$ decreases, generating less efficient structures (\ie, bigger $\tau_c$) which will curtail the growth rate, and vice-versa.  There is a direct analogy to a business enterprise or an army, where an increase in the rate of employment leads to a smaller portion of resources given to optimizing the attachment of any individual new member. 
Thus, during rapid growth spurts, hiring is likely less optimal than during periods of slow growth.  
All techniques used herein are applicable to networks, but for simplicity, we consider a tree  where each arriving node connects to just one parent.  
\begin{figure}[t]
\rotatebox{270}
{\resizebox{!}{2.8in}{\includegraphics{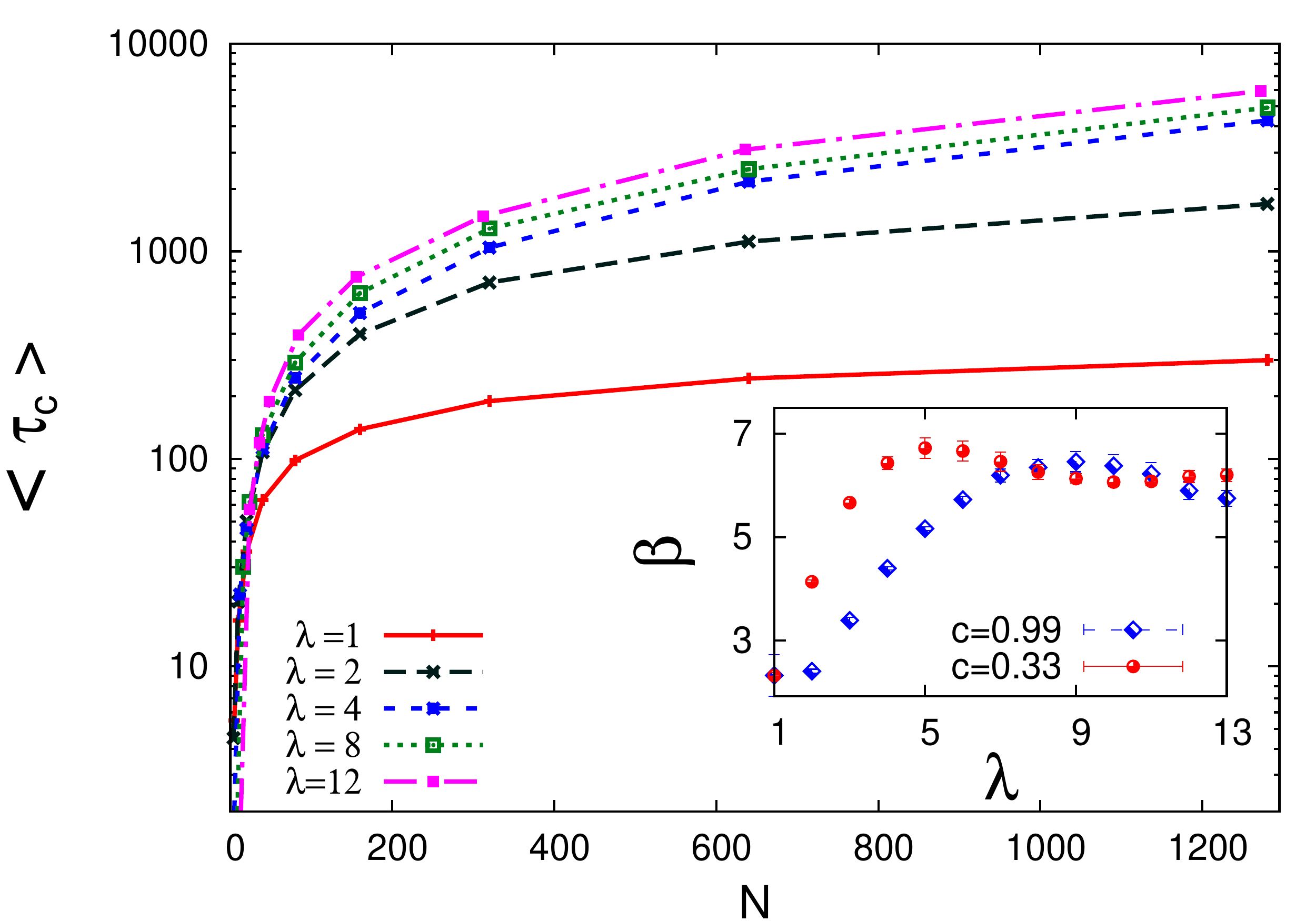}}}
\caption{(Color online) Data points are the expectation value, $\left<\tau_c(N)\right>$, of $1000$ independent realizations with $c=1/3$ and varying values of $\lambda$ as indicated. We find $\left<\tau_c(N)\right> \sim \ln N^\beta$ where $\beta$ depends on $c$ and $\lambda$ as shown in the inset.}
\label{fig:scaling}
\end{figure}

The characteristic time, $\tau_c$, is evaluated as in~\cite{Infocom03} where it was shown to be a performance metric for comparing alternate network topologies. Applications to sensor and to mobile network constructions are discussed in~\cite{Koskinen04,
Badonnel05,Pandey06art} and a similar derivation of $\tau_c$ is  
in~\cite{NohRieger04}. 
A random walk on $G(t)$ is considered, where the walker represents a message to be communicated. We assume unicast communication 
(\ie, a node exchanges messages with only one other node 
at a time) and that all nodes constantly attempt to transmit messages. 
More specifically, if node $i$ is connected to $d_i$ neighbors, it successfully transmits on average $1/d_i$ fraction of the time, to one neighbor chosen at random (\ie, with probability $1/d_i$).  
The remaining $1-1/d_i$ fraction of the time transmission is not successful and the message remains on node $i$, waiting to be transmitted. 
The state transition matrix $P$ describes this process, where element $P_{ij}$ is the probability of the message passing from node $i$ to $j$ at any discrete time step, with $P_{ii}$ the probability of an unsuccessful attempt.  $P_{ij}=0$ if $i$ and $j$ are not directly connected in $G(t)$, otherwise
\begin{equation}
P_{ij} =
\begin{cases}
{{1/d_i}^{2}} & \ \ {\rm if} \ \ i \ne j \\
1-1/d_i & \ \ {\rm if} \ \ i = j. 
\end{cases}
\label{Pmatrix}
\end{equation}
$P$ is column stochastic and irreducible. Let $r_i$ and $\vec{v}_i$ denote the eigenvalues and eigenvectors of $P$. 
By the Perron-Frobenius theorem, there is one eigenvalue \mbox{$r_1=1$} corresponding to the unique steady-state distribution. 
All remaining eigenvalues have $|r_i| < 1$ and are modes that decay to the steady-state. 
The characteristic time $\tau_i$ for mode $i$ to decay by a factor of $1/e$ is defined by the equality $P^{\tau_i}\vec{v}_i = (r_i)^{\tau_i} \vec{v}_i$ and setting  $|r_i|^{\tau_i} = 1 / e$. The longest characteristic time $\tau_c$ results from  
$r_2$ (the largest $r_i < 1$). Rearranging,
$\tau_c = -1/\ln |r_2|.$
 
To implement Eqn.~(\ref{eqn:lambda}) we need to compare ${\cal{F}}(G)$ for two networks with different sizes. Yet as $N$ increases, $\tau_c$ typically increases. No rigorous results exist describing the relationship.
We find empirically that $\left<\tau_c(N)\right> \sim \ln N^\beta$, where $\beta$ depends on $c$ and $\lambda$, as shown in Fig.~\ref{fig:scaling}. Fitness of any particular realization is thus evaluated as ${\cal{F}}(G) = 
-\tau_c/\ln N^\beta$ (relative to the ensemble of networks with those specific values of $c$, $\lambda$ and $N$). Note, the negative sign is due to larger $\tau_c$ being {\it less} fit. 

\begin{figure}[t]
\resizebox{3.in}{!}{\includegraphics{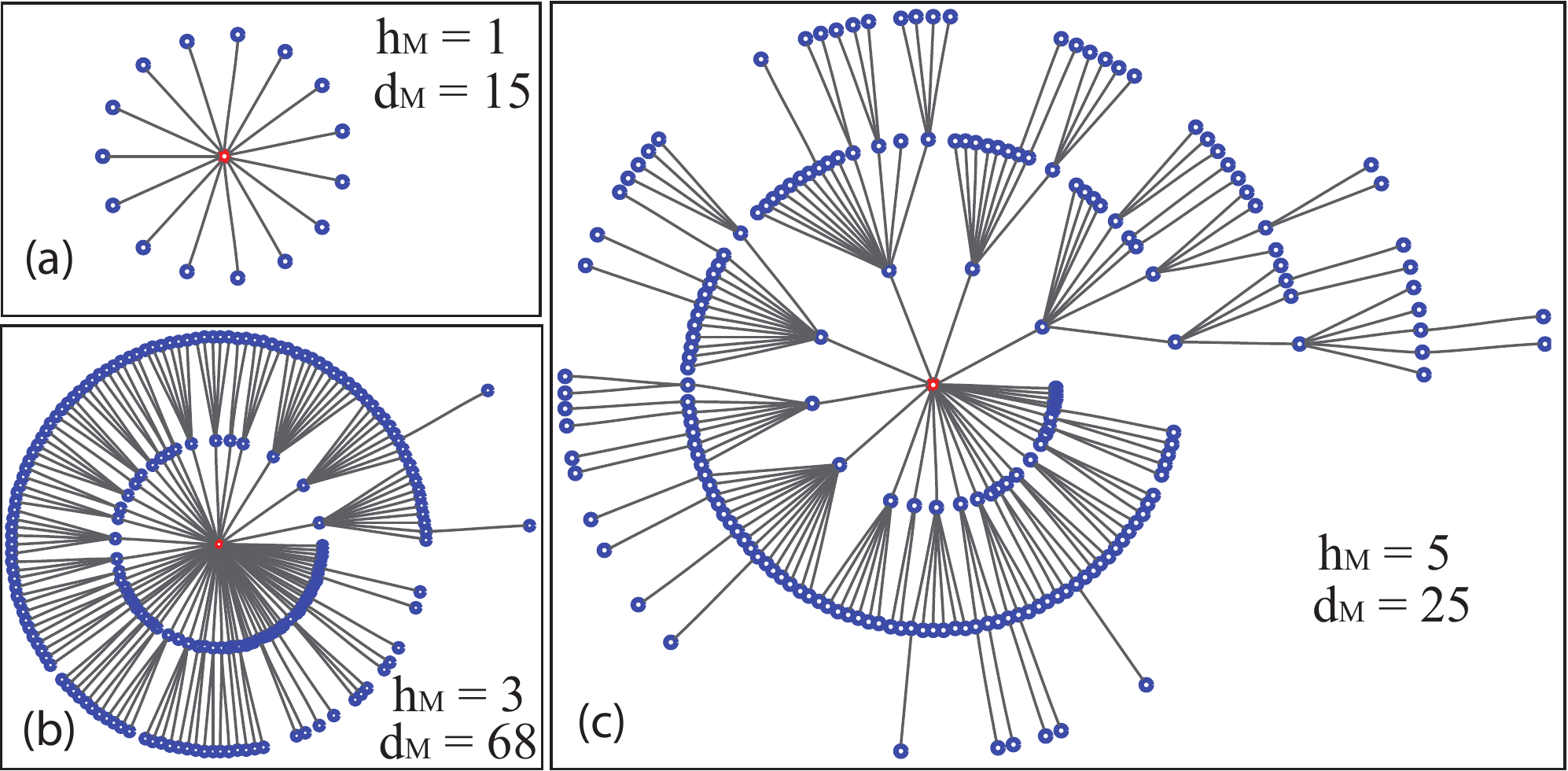}}
\caption{(Color online) 
With no feedback networks range from stars (a) to random recursive trees.  Example of networks with (a) $\frac{c}{\lambda}=1$, $N=16$; (b) $\frac{c}{\lambda} = 1/3$, $N=200$; (c) $\frac{c}{\lambda} = 1/10$, $N=200$, with maximum degree $d_M$ and depth $h_M$ indicated. 
}
\label{fig:vary}
\end{figure}

\begin{figure}[tb]
\resizebox{!}{1.51in}{\includegraphics{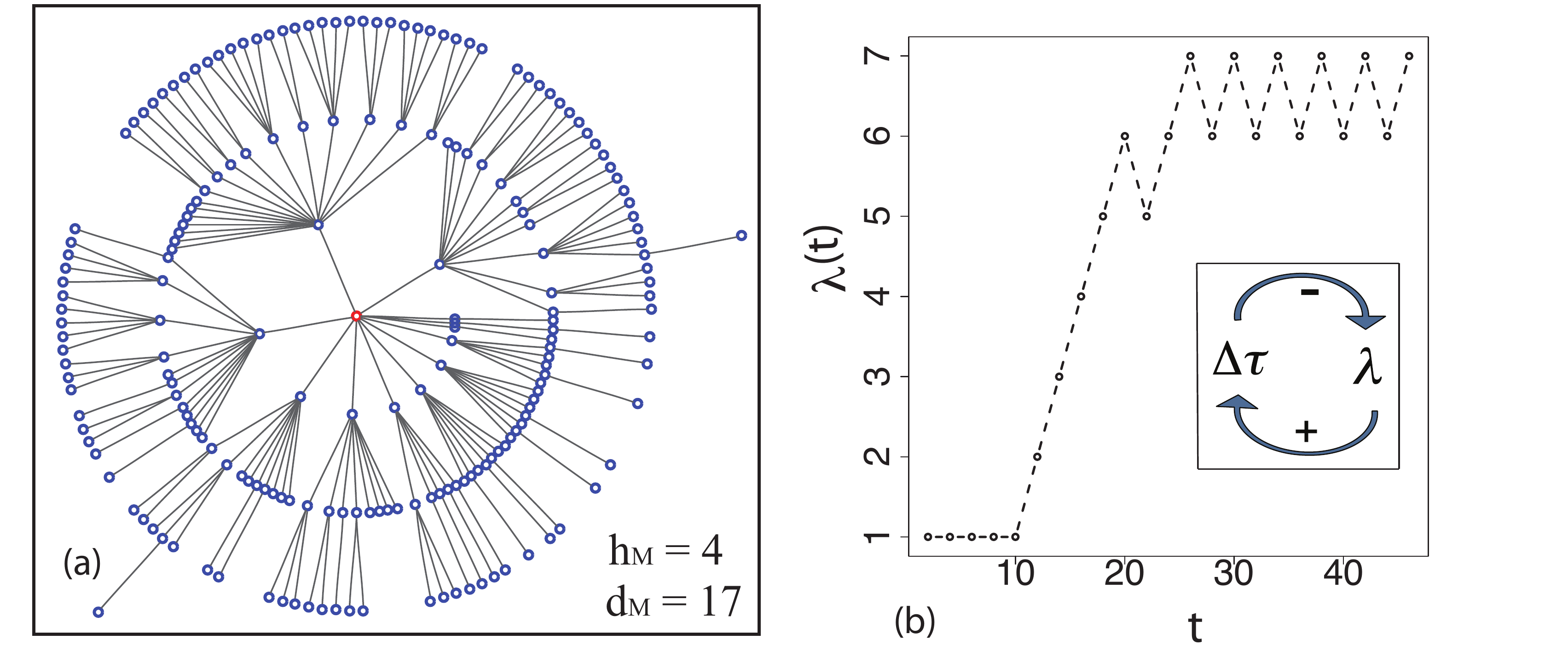}}
\caption{(Color online)  (a) A typical network grown {\it with} feedback.  Here $c=1/3, \lambda_0=1$, $N=200$, and $\delta=2$. This network interpolates dynamically between those in Fig.~\ref{fig:vary}. 
(b) $\lambda(t)$ for this realization. Inset  :  schematic of  feedback loop.}
\label{fig:FBnet}
\end{figure}

We analyze the model above via computer simulation, implementing it in R and visualizing results with Graphviz~\cite{comp}.
First we consider no feedback ($\lambda$ constant)  
and linear scaling of $k(t) = \lceil c N(t)/\lambda \rceil$.
The notation $\lceil a \rceil$ denotes the closest integer greater than or equal to $a$, used since $k$, the number of candidate parents, must be an integer. 
Linear scaling provides intuition on how $c/\lambda$ tunes the structures.
There are two limiting behaviors: $c/\lambda=1$ (\ie, $k=N$) generates a star topology;
and $c/\lambda \rightarrow 0$ (\ie, $k=1$) generates exactly random recursive trees~\cite{SmytheMah95}. Figure~\ref{fig:vary} shows representative networks grown with three different fixed values of $c/\lambda$, with maximum node degree $d_{M}$ and maximum depth $h_{M}$ indicated. 

\begin{figure}[b]
\rotatebox{270}
{\resizebox{1.7in}{!}{\includegraphics{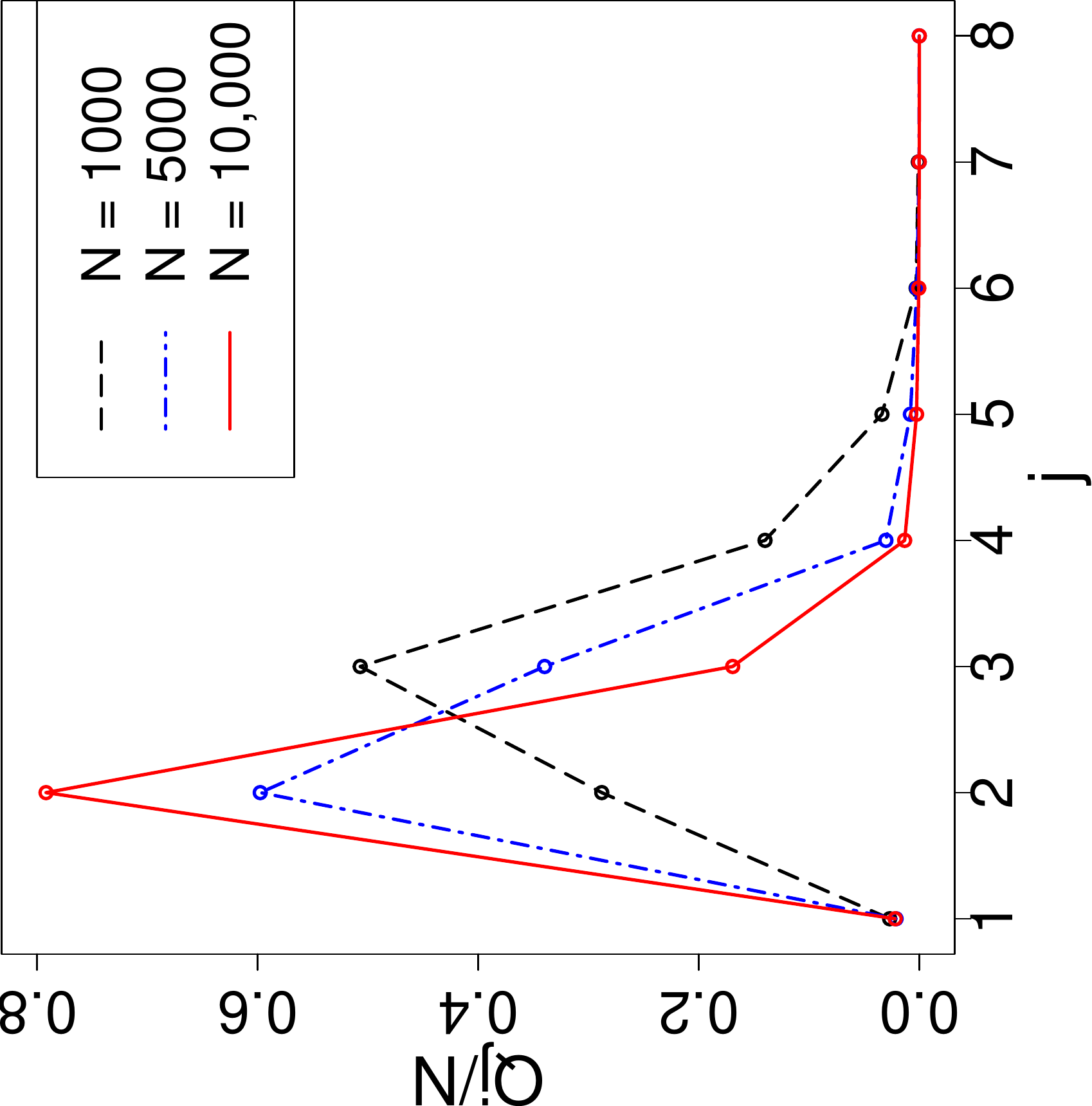}}}%
\hspace*{0.25cm}%
\rotatebox{270}
{\resizebox{1.7in}{!}{\includegraphics{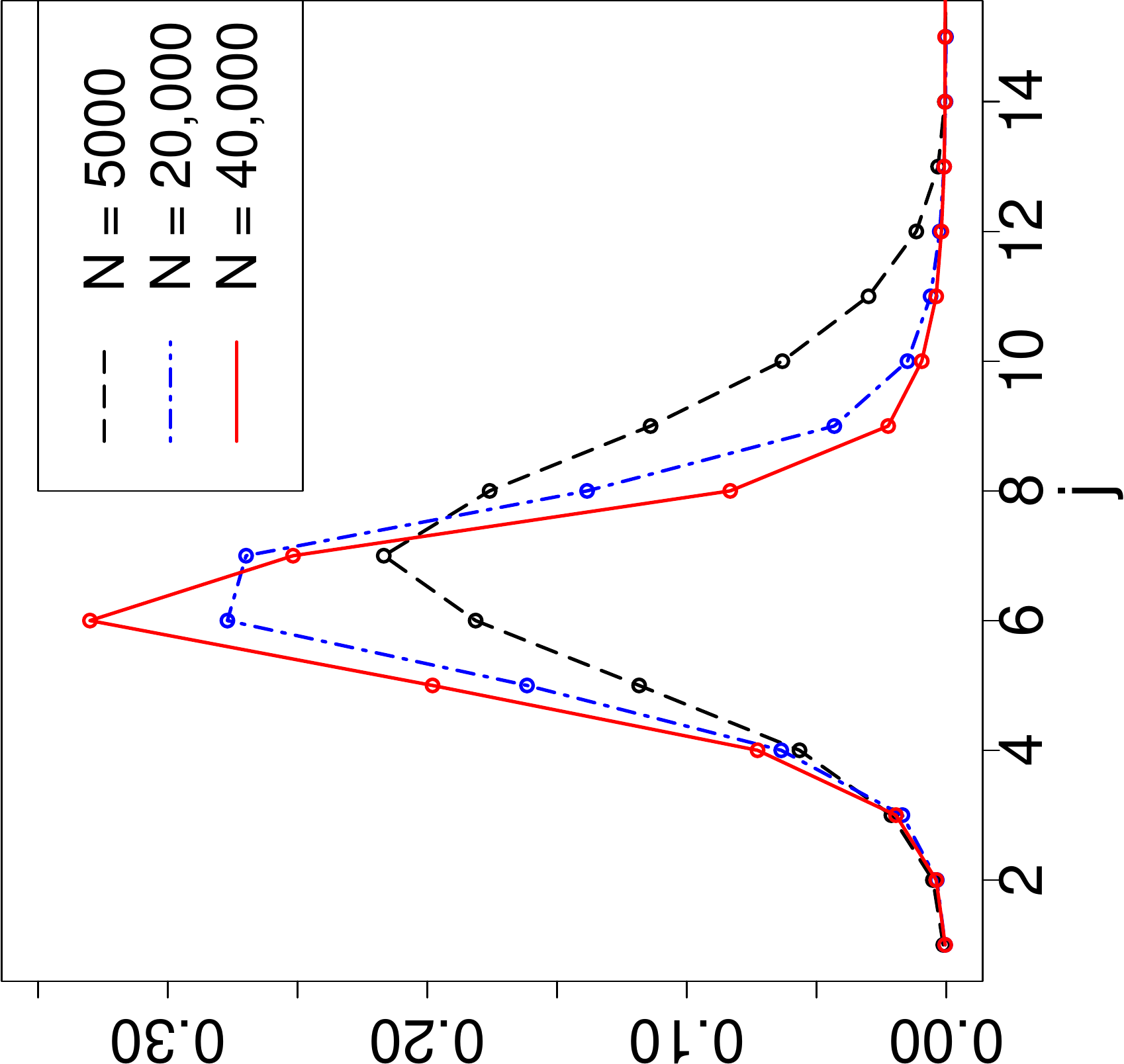}}}%
\caption{(Color online) Depth distribution for $c/\lambda=0.02$. (a) Linear scaling, $k\sim N$, results in condensation to $h \le 2$ for $N > N_x = (\lambda/c)^2$. 
(b) For $k \sim \sqrt{N}$, the distribution remains concentrated well above the condensed structure for $N \ll N_x = 0.25 (\lambda/c)^6$.}
\label{pdfs}
\end{figure}

On incorporating feedback (Eqn.~(\ref{eqn:lambda}) with $\delta$ finite), 
$\lambda$ becomes a tunable parameter. 
For fixed $c$, as $\lambda$ increases the system moves
towards $k=1$ (adding new layers of hierarchy).
As $\lambda$ decreases the system
moves towards $k=N$ (filling in existing levels
of the hierarchy). Thus adjustments in $\lambda$ tune the levels of hierarchy (and the degree assortativity~\cite{MEJN-mixing}). 
Figure~\ref{fig:FBnet}~(a) shows a typical network grown {\it with} feedback where $c=1/3$, $\lambda_0=1$ and $\delta=2$, grown to size $N=200$. It has the same initial conditions and final size as Fig.~\ref{fig:vary}~(b), however, with $d_{M}=17$ it resembles a more balanced version of Fig.~\ref{fig:vary}~(c). Also, the root is no longer the highest degree node. Figure~\ref{fig:FBnet}~(b) shows the evolution of $\lambda(t)$ for this realization which is representative of the typical behavior observed (in particular, the final steady-state oscillation).
For larger networks, the consequences of linear scaling between $k(t)$ and $N(t)$ become manifest even in the absence of feedback. With no feedback, using $k(t) = \lceil c N(t)/\lambda(t) \rceil$ quickly leads to a ``condensed" structure where all new nodes join with depth $h\le 2$.  
Much of the analysis in~\cite{RDPKCM07} applies here, except now $k(t)$ is a function rather than a fixed constant.  
Let $Q_j(t)$ denote the number of nodes at depth $j$, and $\mu_j(t)$ the expected number of nodes at depth $j$ in the candidate set, $\mu_j(t) = k(t) Q_j(t)/N(t)$. 
Boundary conditions are $Q_0(t) = 1$ for all $t \ge 0$, and $Q_j(0)=0$ for all $j \ge 1$. We can explicitly calculate the exact recurrence $Q_1(t) = \sum_{n=1}^{t-1} k(n)/n$.  
Approximating the discrete sum by integration, for $k(t) = cN(t)/\lambda$, we find $Q_1(t) \approx cN(t)/\lambda$ and accordingly, $\mu_1(t) = (c/\lambda)^2 N(t)$.  Thus once $N(t) > N_x = (\lambda/c)^2$, $\mu_1(t) > 1$, and with high probability all further incoming nodes join with $h \le 2$.  Figure~\ref{pdfs}~(a) shows this crossover of the depth distribution for fixed $c/\lambda = 0.02$ (with crossover length $N_x=2500$). 

For sublinear scaling, such as $k(t) = \lceil c \sqrt{N(t)}/\lambda(t) \rceil$,  condensation to $h \le 2$ can be avoided. 
$Q_1(t) \approx 2c\sqrt{N}/\lambda$, and thus $\mu_1 = 2(c/\lambda)^2 < 1$ so long as $c/\lambda < 1/\sqrt{2}$ (independent of $N$).  Yet, we eventually see condensation to depth $h \le 3$  
happen once $Q_2(t)$ grows large. Asymptotically $Q_2(t) = \int \left[k(n) Q_1(n)/n\right] dn \approx 2c^2 N/\lambda^2$.  Once $\mu_2(t) = 2 c^3 N^{1/2}/\lambda^3  > 1$, all subsequent nodes join with $h\le 3$, which occurs at crossover length $N_x = 0.25(\lambda/c)^6$. Figure~\ref{pdfs}~(b) shows the evolution of the depth distribution for $N\ll N_x$ with $c/\lambda = 0.02$ (here $N_x  = 4\times 10^9$).  It becomes more sharply peaked with increasing $N$ and shifts towards lower average depth but remains concentrated well above the final condensed state.
In general we can show that $k\sim N^{1/a}$, for $a$ any integer, ultimately leads to condensation at depths  $h \le a+1$, with crossover as large as 
$N_x \sim (\lambda/c)^{a(a+1)}$~\cite{DKinprep}.
For logarithmic scaling, 
$(k(t) = \lceil c \log (N(t))/\lambda(t) \rceil)$, the peak of the depth distribution increases as $j = \ln N/ \ln \ln N$, and collapse is avoided altogether~\cite{DKinprep}. 

Now that the dependence between $k(t)$ and $N$ is understood in the absence of feedback, we can incorporate feedback.  We are interested in realistic values of $c\sim 1/3$ and networks of $N \sim 1000$ for which square-root scaling, $k(t) = \lceil c \sqrt{N(t)}/\lambda(t) \rceil$, is sufficient to avoid crossover. We numerically generate ensembles of 100 independent realizations at  various values of $c$ and $\lambda_0$, all of which produce similar results. Table~\ref{table} summarizes numerical results for $c=1/3$ and $\lambda_0=3$ (here $N_x \sim 10^5$). Column 2 is the baseline behavior with no feedback. Comparing this with columns 3 and 5 shows that feedback leads to more efficient networks grown to the same size ($N = 501$) in less time, with greater depth and lower maximum degree. Comparing column 2 with 4 and 6 shows that in a given time interval, with feedback, networks grow about twice as large and have improved efficiency. 
In general we find these desirable outcomes are enhanced the more often feedback is evaluated. The time required to attain $N=501$ decreases linearly with decreasing $\delta$ and the network size attained in an allotted time interval increases linearly with decreasing $\delta$.  Of course, each time feedback is assessed requires resources. In our numerical implementation, they are computational resources.  Determining the optimal value of $\delta$ would require assessing the tradeoff between 
this increase in resources and the enhanced network properties. 
For $\delta \lesssim 10$,
our simulations do not  show significant sample-to-sample fluctuations in $\left<\cal{F}(G)\right>$. 
An exhaustive study of self-averaging in networks~\cite{sa-nets} {\it with feedback} may be discussed elsewhere~\cite{DKinprep}.
\begin{table*}[tb]
\begin{tabular}{|l|c|c|c|c|c|}
\hline \ & $\delta \rightarrow \infty$ & $\delta=5$, $N_{{\rm
stop}}=501$ &  $\delta=5$, $t_{{\rm stop}}=167$ &  $\delta=1$, $N_{{\rm stop}}=501$ &  $\delta=1$, $t_{{\rm stop}}=167$\\
\hline
$\left<N(t)\right>$ & 501 & 501 & $956 \pm 11$ & 501 & $1101 \pm 7 $\\
$\left<{\rm time}\right>$ & 167 & $89.3 \pm 0.5$ & 167 & $77.1 \pm 0.6$ & 167\\
$\left<h\right>$ & $3.08  \pm 0.02$ & $3.41  \pm 0.03$ & $3.72  \pm 0.03$ & $3.38 \pm 0.02$ & $3.76 \pm 0.02$\\
$\left<h_{\rm M} \right>$ & $6.3 \pm 0.7$ & $7.09 \pm 0.08$ & $7.72 \pm 0.08$ & $7.10 \pm 0.08$ & $7.82  \pm 0.09$\\
$\left<d_{\rm M} \right>$ & $32.2 \pm 0.6$ & $25.2 \pm 0.4$ & $31.1 \pm 0.4$ & $24.8 \pm 0.4$ & $32.7 \pm 0.5$\\
$< \tau_c/\ln N^\beta>\ $ & \ $0.057 \pm 0.001$ \  & $0.014  \pm 0.001$ & $0.014  \pm 0.001$ &  $ 0.014 \pm 0.001$ & $0.016 \pm 0.001$ \\ \hline
\end{tabular}
\caption{Average network properties, over 100 independent realizations, for $c=1/3$, $\lambda_0=3$, and $k(t) = \lceil c \sqrt{N(t)}/\lambda(t) \rceil$. 
Comparing columns 2, 3, and 5 shows that with feedback networks grow to be of size $N=501$ in much shorter time and are more efficient (smaller $\tau_c/\ln N^{\beta}$). They also have greater maximum depth and lower maximum degree. Comparing columns 2, 4 and 6 shows that in a given time interval, networks with feedback grow to be to about twice the size and are more efficient.}  
\label{table}
\end{table*}
In summary,  we introduce a general framework for incorporating feedback into network growth models. Proof of concept is demonstrated using a simple model of a hierarchical network where limited resources are shared amongst all arriving nodes, vying to minimize their distance to the root. Feedback leads to growth of larger, more efficient structures. Linear scaling of resources results in crossover to a trivial condensed structure which can be considerably delayed with sublinear scaling. 
In the context of a growing organization, this suggests 
sublinear scaling is necessary once $N \sim 1000$. 
It may be possible to obtain rigorous results for this model of network growth with feedback by interpreting $c/\lambda$ as a branching rate 
or by proving convergence of $\lambda(t)$ to steady-state oscillation. 

The general framework  proposed herein allows flexibility in choosing other growth models, communication models between nodes $P_{ij}$ (\eg, broadcast rather than unicast), and fitness functions ${\cal F}(G(t))$ (\eg, fitness landscapes~\cite{Mitchell96} 
modeling random evolutionary pressures). 
Such alternate choices of ${\cal F}(G(t))$ may overcome the current limitation that global topology information is required to assess $\tau_c$.  An alternate growth model, where nodes maximize their distance to the root, also seems to demonstrate similar effects of  feedback

Other recent models that could provide mechanisms for introducing feedback are the TARL model of two interacting networks~\cite{BornerPNAS04}, a generative model where loops within a network are considered as potential feedback channels~\cite{WhiteTsallisFarmer06}, and the layered network framework of ~\cite{KurantThiranPRL06}.

Discussions with C. D'Souza, 
J. Chayes, C. Borgs, P. Krapivsky, J. Machta
and C. Moore, and attendance at MSRI's  ``Real World Data Networks" and IPAM's ``Random Shapes" workshops are gratefully acknowledged.
\bibliographystyle{unsrt}

\end{document}